\documentclass[%
 reprint,
superscriptaddress,
 amsmath,amssymb,
 aps,
]{revtex4-2}

\usepackage{booktabs}
\usepackage{graphicx}% Include figure files
\usepackage{xcolor}
\usepackage{dcolumn}% Align table columns on decimal point
\usepackage{bm}% bold math
\usepackage{multirow} 
\setcitestyle{super} 

\usepackage{array}

\newcolumntype{L}[1]{>{\raggedright\let\newline\\\arraybackslash\hspace{0pt}}m{#1}}
\newcolumntype{C}[1]{>{\centering\let\newline\\\arraybackslash\hspace{0pt}}m{#1}}
\newcolumntype{R}[1]{>{\raggedleft\let\newline\\\arraybackslash\hspace{0pt}}m{#1}}

\newcommand*{\citen}[1]{%
  \begingroup
    \romannumeral-`\x % remove space at the beginning of \setcitestyle
    \setcitestyle{numbers}%
    \cite{#1}%
  \endgroup   
}

\begin{document}

\preprint{APS/123-QED}

\title{All-Electron Relativistic Fully Self-Consistent $GW$ Study of Heteronuclear Actinide-Containing Diatomics}

\author{Jacob Adamski}
\affiliation{%
 Department of Chemistry, University of Michigan, Ann Arbor, MI 48109.
}% 
\author{Vibin Abraham}
\email{avibin@umich.edu}
\affiliation{%
 Department of Chemistry, University of Michigan, Ann Arbor, MI 48109.
}% 
\affiliation{Current address: Physical and Computational Science Directorate, Pacific Northwest National Laboratory, Richland, Washington 99354, USA}

\author{Dominika Zgid}
\email{zgid@umich.edu}
\affiliation{%
 Department of Chemistry, University of Michigan, Ann Arbor, MI 48109.
}%
\affiliation{%
 Department of Physics and Astronomy, University of Michigan, Ann Arbor, MI 48109.
}%
\affiliation{%
Faculty of Physics, University of Warsaw, 02-093 Warsaw, Poland
}%

\date{\today}

\begin{abstract}
The fully self-consistent $GW$ (sc$GW$) approximation provides a Green's-function approach that is starting-point independent and offers a favorable cost-to-accuracy balance compared to high-level wavefunction methods.
Here, we present an all-electron sc$GW$ study of uranium-containing diatomics (UC, UN, UO, and UF), incorporating relativistic effects through the exact two-component (X2C) formalism. 
We evaluate adiabatic ionization energies as well as electron-attachment and detachment energetics (AEA and VDE), together with equilibrium structures and harmonic vibrational frequencies, and we assess their sensitivity to basis-set choice and relativistic treatment.
We find that sc$GW$ yields ionization energies and vibrational properties in very good agreement with experiment and high-accuracy theoretical estimates. For AEA and VDE, diffuse basis sets are essential for convergence.  UF is a particularly challenging case for scalar relativistic methods because its electron-attachment and  vertical detachment energies are strongly affected by spin--orbit coupling, highlighting the need for a variational two-component treatment.
These results establish all-electron X2C-sc$GW$ as a practical route for accurate actinide-molecule energetics and spectroscopy and motivate future applications to larger uranium-containing systems.
\end{abstract}

\maketitle

\section{\label{sec:intro}Introduction}

Traditional quantum chemistry methods, such as coupled cluster (CC) and complete active space second-order perturbation theory (CASPT2), are well known for their ability to describe electron correlation and relativistic effects accurately in molecular systems. However, their steep computational cost makes them impractical for routine applications to solids. In practice, density functional theory (DFT), including hybrid and DFT+$U$ variants, is often the only feasible option for larger actinide systems~\cite{oakley2024trends,duignan2017assessment,lan2015quasi,wittmann2024extension,south20164,batista2015computational}. Yet for compounds with partially occupied $5f$ shells, DFT predictions can depend strongly on the chosen functional and on the degree of electron localization, motivating the search for more systematically improvable many-body approaches.

In this context, the $GW$ method~\cite{Hedin1965,Onida2002,Golze2019,Brunevalgwreview} offers a promising middle ground. Widely used in the physics and materials science communities, it has a more favorable computational scaling than traditional high-level wavefunction methods or Green's function methods containing exchange for both molecular and solid-state systems~\cite{Foerster2011,Ren2012,Krause2015,Caruso2016,Mejia2021,Duchemin2020,kaplan2016quasi,van2018,Golze2020,Galleni2022,Li2022,Mejia2022,Iskakov2019}. At the same time, recent developments have significantly improved both its accuracy and efficiency. With resolution-of-identity techniques, finite-temperature $GW$ can be implemented with $\mathcal{O}(n^4)$ scaling\cite{Yeh2022,Iskakov2020,Shee_GW_SEET}, and tensor hypercontraction can reduce this further to $\mathcal{O}(n^3)$~\cite{Yeh2023_THC_RPA,Yeh2024_THC_GW}. Moreover, fully self-consistent $GW$ (sc$GW$) schemes enabled by modern imaginary-time grids~\cite{Shinaoka2017} remove the reference-state dependence of $G_0W_0$ approaches~\cite{Stan2009,Rostgaard2010,Caruso2012,Yeh2022,Grumet2018,Strange2011,Caruso2013,Pokhilko:local_correlators:2021}, while allowing consistent evaluation of total energies, ionization potentials (IP), and electron affinities (EA). Moreover, if improved accuracy is desired, the $GW$ method can also be used as a first step in vertex-correct $GW$  ($GW\Gamma$) schemes~\cite{wen2023comparing,Pokhilko2024THCGF2GWSOX,Pokhilko2025VertexCorrectedGW, Forster2024WhyGWAccurate,Forster2023,Forster2025VertexCorrections,Bruneval2024G3W2} or $GW$-based embedding methods~\cite{Zgid2017SEET,Shee_GW_SEET,Werner_multitier,Bierman_GW+DMFT_2003,Sheng2022QDET}.

Among the most demanding tests for any electronic structure method are systems containing heavy elements, where relativistic and correlation effects must be treated on equal footing. Although approximate relativistic corrections have been incorporated into a variety of $GW$ implementations for molecules and solids~\cite{Sakuma2011,Kutepov2012,Molina2013,Irene2013,umari2014relativistic,Kuhn2015,Scherpelz2016,Holzer2019,Forster2023}, applications to actinides remain comparatively limited and have focused mainly on solids~\cite{ahmed2014gw,jiang2010first,Kutepov2012} with large pseudo-potentials. As a result, the accuracy of $GW$ for molecular compounds containing very heavy atoms is still not well established.

Understanding the electronic structure of small uranium-containing molecules is important for both theoretical modeling and experimental spectroscopy~\cite{goncharov_probing_2006,martin_ab_1996,paulovic}. Uranium diatomics with second-row ligands are especially appealing benchmark systems because they exhibit multiple oxidation states, unusual bonding patterns, open-shell $5f$ and $6d$ manifolds, and strong relativistic effects~\cite{kovacs2015quantum,AUTSCHBACH2024177,kaltsoyannis2024understanding,dolg2015computational,zhang_route_2022,romeu_energetic_2024,de_melo_electronic_2022,de_melo_bonding_2023,de_melo_theoretical_2022,yousfi2023theoretical,infante2010ionization,antonov_spectroscopic_2013,batista2004density,AnhHighRes2023,matthew_resonant_2013,battey_spectroscopic_2020}. At the same time, their small size makes them accessible to high-level reference calculations, providing a stringent setting for method assessment. 

Historically, accurate modeling of such systems has relied heavily on multireference wavefunction techniques, particularly CASSCF and CASPT2, including spin--orbit extensions such as SO-CASPT2~\cite{ning2023chemical,kovacs2017relativistic,north2023multireference,kervazo2019accurate,gendron2017puzzling,sarkar2023multiconfiguration,zaitsevskii2023theoretical,roos2004relativistic,yousfi2023theoretical,battey_spectroscopic_2020}. These methods are well suited to the near-degeneracies and strong spin--orbit effects that often arise in actinide chemistry. Benchmark-quality energetics for heavy-element systems have also been achieved with Fock-space coupled-cluster (FSCC) methods~\cite{Saetgaraev2025,Kaygorodov2021}. 

For many uranium diatomics, however, the low-lying states of interest show substantial single-reference character, as indicated by both determinant- and spinor-based diagnostics~\cite{infante2010ionization,feng2021coupled}. This makes high-level coupled-cluster methods reliable benchmarks for ground-state properties and ionization potentials~\cite{vasiliu2020calculated,andriola2023coupled,jackson2008prediction}. Composite coupled-cluster strategies that combine accurate correlation treatments with scalar-relativistic and spin--orbit corrections, most notably within the Feller--Peterson--Dixon (FPD) framework, have therefore proven especially useful for uranium diatomics~\cite{dixon2012practical,peterson2012chemical,feller2012further,north2022ab,de_melo_theoretical_2022,de_melo_electronic_2022,romeu_energetic_2024}. 

In this work, we assess the performance of a fully relativistic Green's-function approach for the uranium diatomics UC, UN, UO, and UF. Specifically, we use our recently introduced two-component self-consistent $GW$ method (X2C-sc$GW$) for molecular and periodic systems~\cite{Yeh2022x2c,abraham_relativistic_2024,Harsha_Challenges} to compute adiabatic ionization energies, electron affinities, vertical detachment energies, equilibrium bond lengths, and harmonic vibrational frequencies. By comparing against high-level wavefunction benchmarks and examining basis-set requirements together with the role of variational spin--orbit coupling, we establish the accuracy of X2C-sc$GW$ for actinide molecules and provide guidance for its application to larger uranium-containing systems and materials.

The remainder of this paper is organized as follows: In Section II we briefly outline the sc$GW$ method and the two-component relativistic Hamiltonian employed in our X2C formulation. 
Section III details the computational protocol, basis sets, and convergence criteria. Section IV presents and discusses our results for adiabatic IP, adiabatic EA, vertical detachment energy and vibrational properties, analyzing the performance of sc$GW$ relative to benchmark data. Section V summarizes our main findings and provides an outlook for future applications of relativistic sc$GW$ to larger actinide-containing systems and extended materials.

\section{Methods}\label{sec:methods}

We employ the sc$GW$ method, incorporating relativistic effects through the X2C framework.~\cite{kutzelnigg_quasirelativistic_2005,liu_quasirelativistic_2007,liu_exact_2009,sun_exact_2009,cheng_analytic_2011} While comprehensive details regarding the theory and implementation are available in Refs.~\citen{Yeh2022x2c,Yeh2022,abraham_relativistic_2024}, this work provides a concise summary of the sc$GW$ approach utilized.

%\subsection{self-consistent $GW$}

We work with the imaginary-axis (finite-temperature or Matsubara imaginary axis) formalism described in Refs.~\citen{Yeh2022x2c,Yeh2022,abraham_relativistic_2024,Harsha_Challenges}.
The one-particle imaginary-time Green's function is defined as
\begin{equation}
G_{p q}(\tau)=-\frac{1}{\mathcal{Z}} \operatorname{Tr}\left[e^{-(\beta-\tau)(\hat{H}-\mu \hat{N})} a_p e^{-\tau(\hat{H}-\mu \hat{N})} a_q^{\dagger}\right],
\end{equation}
where $\mathcal{Z} = \mathrm{Tr} \left(
        e^{-\beta (H - \mu N)}
    \right)$ is the partition function, $\beta = 1 / k_B T$ is the inverse temperature ($k_B$ is the Boltzmann constant), $\mu$ is the chemical potential, $\hat{H}$ and $\hat{N}$ are the Hamiltonian and the number operators, $a_{p}$ and $a_{p}^{\dagger}$ are the annihilation and creation operators and $\tau \in \left[0, \beta\right)$ is the imaginary time.

The self-energy in the $GW$ approximation is given as
\begin{equation}
    \mathbf{\Sigma} (i\omega_n)
    = \mathbf{\Sigma}_\infty 
    + \mathbf{\Sigma}^c (i \omega_n),
    \label{eq:self-energy-v1}
\end{equation}
where $\mathbf{\Sigma}_\infty$ is the static, frequency-independent self-energy evaluated in the same way as Hatree-Fock self-energy but using the correlated density matrix.  The dynamic self-energy contribution, $\mathbf{\Sigma}^c (i\omega_n)$, is defined on the imaginary time ($\tau$) axis as
\begin{align}
    \Sigma^c_{pq} (\tau) = - \sum_{ab}G_{ab} (\tau) \tilde{W}_{pabq} (-\tau),
\end{align}
where $\tilde{W}$ is the screened Coulomb interaction.
The self-energy at the $GW$ approximation describes electronic correlations by summing over all the bubble diagrams.
The new Green's function is then defined by the Dyson equation,
\begin{equation}
    \left[ \mathbf{G} (i \omega) \right]^{-1}
    = (i\omega + \mu) \mathbf{S} - \mathbf{H}_0 - \mathbf{\Sigma}^c (i \omega),
    \label{eq:dyson}
\end{equation}
where the chemical potential $\mu$ is fixed to ensure a correct particle number, $\mathbf{S}$ is the overlap matrix, and $\mathbf{H}_0$ is the one-electron Hamiltonian.
The non-interacting Hamiltonian $\mathbf{H}_0$ is formed using the exact two-component (X2C) transformation of the four-component Dirac Hamiltonian.~\cite{liu_exact_2009,cheng_analytic_2011,sun_exact_2009}

The total electronic energy is calculated using the Galitskii--Migdal
formula~\cite{Holm2000}
\begin{subequations}
    \begin{align}
        E_\mathrm{Total}
        &=
        E_\mathrm{nuclear}
        +
        E_{1b}
        +
        E_{2b},
        \\
        E_{1b}
        &=
        \mathrm{Re}\,
        \left\{
        \mathrm{Tr}
        \left[
            \rho \mathbf{H}_{0}
        \right]
        +
        \frac{1}{2}
        \mathrm{Tr}
        \left[
            \rho \mathbf{\Sigma}_{\infty}
        \right]
        \right\},
        \\
        E_{2b}
        &=
        \frac{1}{2\beta}
        \sum_{n=-\infty}^{\infty}
        e^{i\omega_n 0^+}
        \mathrm{Tr}
        \left[
            \mathbf{G}(i\omega_n)
            \mathbf{\Sigma}^c(i\omega_n)
        \right],
    \end{align}
\end{subequations}
where $E_\mathrm{nuclear}$ is the nuclear repulsion energy,
$\rho=-G(\tau=\beta^-)$ is the one-particle density matrix, and
$\mathbf{\Sigma}_{\infty}$ is the static Hartree--Fock-like part of the
self-energy. 

Relativistic effects are treated within the X2C framework, which decouples the four-component one-electron Dirac Hamiltonian into an electrons-only form.~\cite{kutzelnigg_quasirelativistic_2005,liu_quasirelativistic_2007,liu_exact_2009,sun_exact_2009,cheng_analytic_2011} In this work, both the spin-free X2C (sfX2C) and full X2C Hamiltonians are employed in the sc$GW$ calculations. The former accounts only for scalar relativistic effects, while the latter additionally includes spin--orbit coupling.

\section{\label{sec:comp_detail} Computational Detail}

We used the correlation-consistent basis sets developed by Peterson et al., specifically cc-pvXZ-DKH3 (X = T, Q), parameterized for DKH3 corrections for the uranium (U) atom. This basis is commonly used with X2C methods.~\cite{peterson_correlation_2015}
For the second row atoms (C, N, O, and F), we used Dunning's cc-pvXZ (X = T, Q) basis sets.~\cite{dunning_gaussian_1989}
The mean-field HF calculations and the density-fitted integrals were obtained using PySCF 2.2.1.\cite{sun_recent_2020}
For electron affinities and vertical detachment energies, we included diffuse functions on the ligand atoms by using Dunning’s augmented correlation-consistent basis sets, aug-cc-pVXZ (X = T, Q).

We used the experimental geometries whenever possible. When such geometries were not available, we employed the geometries evaluated with the SO-CASPT2  method.\cite{battey_spectroscopic_2020, de_melo_theoretical_2022, de_melo_electronic_2022, kaledin_laser_1994, romeu_energetic_2024, bross_theoretical_2015} 
The geometries of the neutral, cationic and anionic systems are listed in the supporting information.

The sc$GW$ calculations were carried out at $\beta = 1000$ [a.u.$^{-1}$]. 
We employ intermediate representation (IR) grids to express the dynamic quantities in this work which utilize sparse sampling along the imaginary-time and Matsubara frequency axes.~\cite{Shinaoka2017,Li2020}. 
We use a development version of the code \texttt{green-mbpt}~\cite{green_mbpt,iskakov_greenweakcoupling_2025} for all the $GW$ calculations.
We employed the direct inversion of iterative subspace (DIIS) to accelerate the  convergence of the sc$GW$ iterations; however, it was modified to use the commutator of the Fock matrix and the single particle density matrix as the error vector.~\cite{Pokhilko2022}

The total energy was extrapolated to the CBS limit using a two point extrapolation scheme with the energies from the  TZ and QZ basis sets \cite{martin_ab_1996}:
\[
E = E_\text{CBS} + \frac{A}{(n + 1/2)^4},
\]
where $n$ is the cardinal number of the basis set, $A$ is a fitting parameter, and $E_\text{CBS}$ is the estimated CBS energy.

In the Supporting Information, we provide the effect of two-electron relativistic
corrections on the sc$GW$ IPs and EAs at the triple-$\zeta$ level, the potential energy
curves for the neutral species, and the effect of spin--orbit coupling on the adiabatic
electron affinities of UC, UN, and UO.

\begin{table*}[t]
\centering
\caption{Spin-resolved Mulliken-type AO populations obtained from the correlated one-particle density matrix for the U atom and ligand atom at the X2C level of relativistic correction for sc$GW$.}
\label{tab:no_analysis}
\setlength{\tabcolsep}{5pt}
\renewcommand{\arraystretch}{1.2}
\begin{tabular}{l l rrr rrr rrr rrr}
\toprule
\multirow{2}{*}{Atom} & \multirow{2}{*}{Orbital} 
 & \multicolumn{3}{c}{UC} & \multicolumn{3}{c}{UN} & \multicolumn{3}{c}{UO} & \multicolumn{3}{c}{UF} \\
\cmidrule(lr){3-5} \cmidrule(lr){6-8} \cmidrule(lr){9-11} \cmidrule(lr){12-14}
 & & $\alpha$ & $\beta$ & Total & $\alpha$ & $\beta$ & Total & $\alpha$ & $\beta$ & Total & $\alpha$ & $\beta$ & Total \\
\midrule
\multirow{3}{*}{U} 
 & 5f & 2.36 & 0.50 & 2.85 & 2.29 & 0.59 & 2.88 & 2.73 & 0.47 & 3.20 & 2.66 & 0.42 & 3.08 \\
 & 6d & 1.06 & 0.85 & 1.91 & 0.93 & 0.80 & 1.73 & 0.80 & 0.49 & 1.28 & 0.33 & 0.28 & 0.61 \\
 & 7s & 0.90 & 0.11 & 1.01 & 0.90 & 0.07 & 0.98 & 0.92 & 0.05 & 0.97 & 0.93 & 0.92 & 1.85 \\
\midrule
\multirow{2}{*}{Ligand} 
 & 2s & 0.89 & 0.84 & 1.73 & 0.90 & 0.90 & 1.80 & 0.94 & 0.95 & 1.89 & 0.99 & 0.99 & 1.98 \\
 & 2p & 1.54 & 1.14 & 2.68 & 1.70 & 1.87 & 3.57 & 2.23 & 2.34 & 4.56 & 2.70 & 2.71 & 5.41 \\
\midrule
\multicolumn{2}{l}{U valence config.}
 & \multicolumn{3}{c}{$5f^2 7s^1$}
 & \multicolumn{3}{c}{$5f^2 7s^1$}
 & \multicolumn{3}{c}{$5f^3 7s^1$}
 & \multicolumn{3}{c}{$5f^3 7s^2$} \\
 \bottomrule
\end{tabular}
\end{table*}

\section{\label{sec:res} Results}

In this section, we present a comprehensive assessment of the electronic and spectroscopic properties of the uranium monochalcogenides, mononitrides, monoxides, and monofluorides. Our analysis focuses on three central aspects: (i) the behavior and convergence of the sc$GW$ procedure for actinide-containing molecules, including the role of relativistic effects and starting references; (ii) the electronic structure of the targeted ground states, characterized through natural orbital occupations; and (iii) adiabatic and vertical ionization and electron-attachment energies, along with equilibrium bond lengths and vibrational frequencies. By comparing sfX2C and X2C results, analyzing basis-set effects, and benchmarking against high-accuracy theory and experiment, we evaluate the performance and limitations of sc$GW$ for these heavy-element systems.

\begin{figure}
  \centering
  \includegraphics[width=\linewidth]{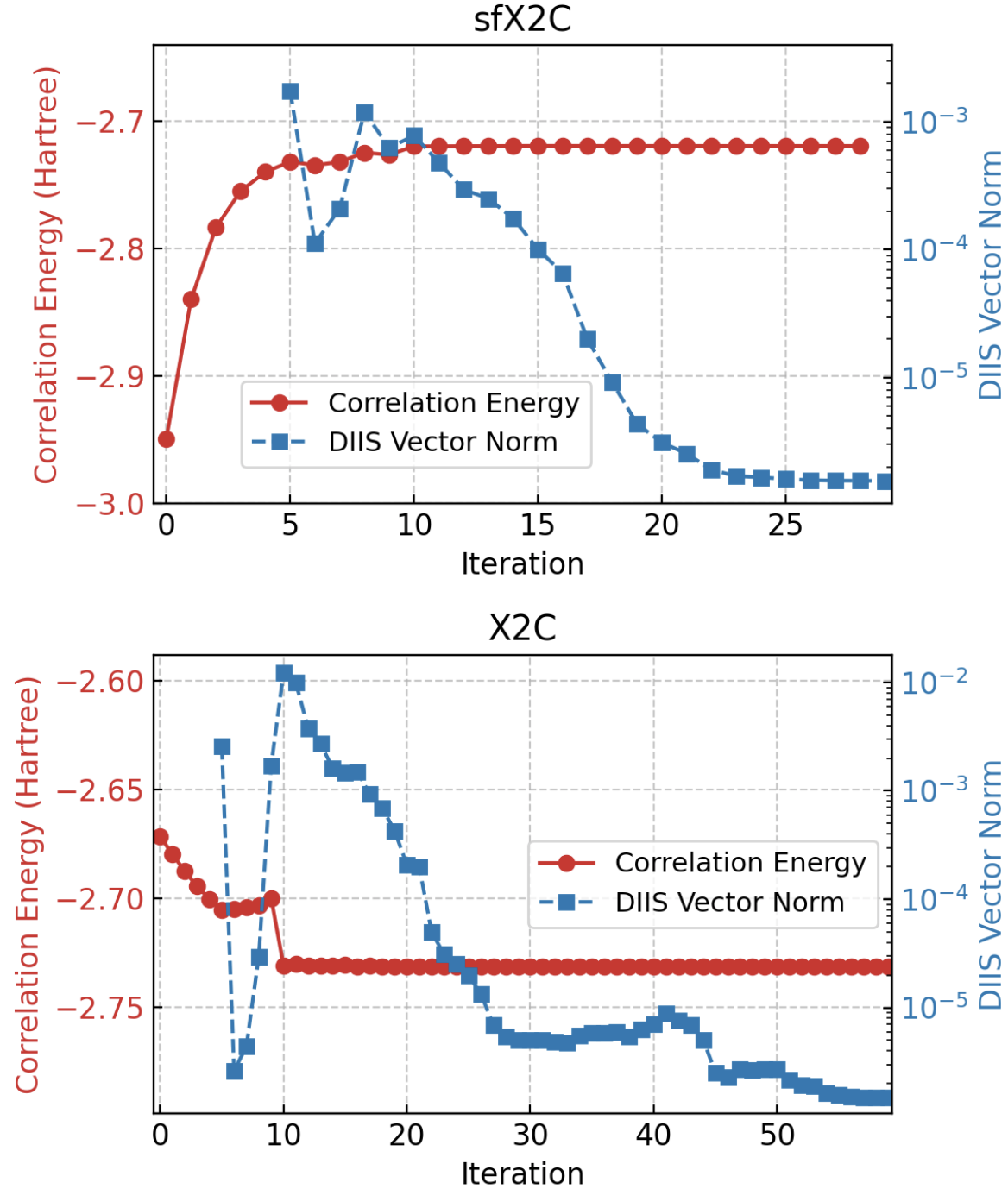}
  \caption{Convergence of sc$GW$ iterations for the UN system. Top: sfX2C starting from PBE0. Bottom: X2C starting from HF reference. The DIIS algorithm is used after the 5th iteration. }
  \label{fig:un}
\end{figure}
\subsection{Obtaining sc$GW$ solution}

Achieving self-consistency and convergence to a desired solution is a known challenge for all iterative methods involving a set of non-linear equations, including sc$GW$.\cite{Caruso2014,Pokhilko2022,Pokhilko2025HomotopyDyson} These calculations can be slow, requiring many iterations and significantly increasing the overall computational cost. Here, we discuss how self-consistency was achieved for the uranium nitride (UN) compound using the commutator DIIS procedure with only the static quantities.\cite{Pokhilko2022} 
The chemical potential was reoptimized at every iteration to ensure that the number of electrons remained fixed.

For the sfX2C method, we used a starting guess from the hybrid functional PBE0 to obtain a reliable initial guess. We used a damping factor of $\alpha = 0.5$ for the first five iterations in the sfX2C calculations. 
When the damping factor is used, the resulting static self-energy is updated as
\begin{equation}
\Sigma = (1-\alpha)\Sigma_{old} + \alpha\Sigma_{new}.
\end{equation}
A characteristic feature of the convergence profile (top panel of Fig. \ref{fig:un}) is the sudden jump in the DIIS vector norm during the early iterations; this behavior reflects a reoptimization of the orbitals. After this regime, the CDIIS procedure stabilizes the iterations, and the total energy converges smoothly to within at least $10^{-5}$ a.u.

For the X2C method we used the generalized Hartree Fock (GHF) as the starting point. We used a damping factor of $\alpha = 0.3$ for the first five iterations in the X2C calculations. 
As in the sfX2C case, the early-stage spike in the DIIS vector norm and the corresponding shift in the correlation energy arise from orbital reoptimization, which is more pronounced here due to the additional degrees of freedom in the GHF reference. Once CDIIS is applied, the iterations gradually stabilize and the total energy converges to at least $10^{-5}$ a.u. We also list the DIIS vector norm at each iteration $i$, which is defined as $\mathbf{e}_i = \left[ F_i, P_i \right]$. Similar convergence profiles were also observed in previous studies.~\cite{Pokhilko2022}

\subsection{Electronic structure analysis of the targeted states}
To characterize the ground-state electronic structure obtained from sc$GW$, we analyze the converged one-particle density matrix together with the corresponding natural orbitals. The atomic populations reported in Table~\ref{tab:no_analysis} were obtained from the correlated one-particle density matrix in the atomic-orbital basis using a Mulliken-type population analysis, i.e., from the $D\cdot S$ population matrix, where $D$ is the AO density matrix and $S$ is the AO overlap matrix. Populations for the U $5f$, $6d$, and $7s$ manifolds and the ligand $2s$ and $2p$ manifolds were then formed by summing contributions over the corresponding AO subspaces. The assignments discussed below are based on these AO-resolved populations and on inspection of the leading natural orbitals and their occupancies, obtained by diagonalizing the converged one-particle density matrix. We carry out this analysis for the ground state neutral configuration for these molecules.
Note that the configurations discussed below refer to the dominant uranium-centered valence pattern. Because the correlated density is delocalized over uranium and ligand centers, the Mulliken populations in Table~\ref{tab:no_analysis} are noninteger and are interpreted together with the leading natural orbitals and their occupancies.

With this interpretation, the sc$GW$ natural orbitals support a consistent picture across the series. For UC and UN, the dominant uranium-centered configuration is $5f^27s^1$, in agreement with earlier high-level studies. In both molecules, the approximately one-electron U $7s$ population in Table~\ref{tab:no_analysis} is consistent with a singly occupied uranium $7s$ orbital, while the additional U $5f$ and $6d$ populations reflect participation of these shells in the U--ligand bonding manifold rather than a pure atomic $5f^3$ assignment. For UC, this is especially clear because the ground state contains, in addition to the uranium-centered $5f^27s^1$ pattern, a singly occupied U--C $\sigma$ bonding orbital with mixed ligand $2p$ and U $5f/6d$ character.

For UO, the natural orbital analysis is consistent with a dominant uranium-centered $5f^37s^1$ configuration, while UF is best described as $5f^37s^2$. The U $7s$ populations in Table~\ref{tab:no_analysis} support this distinction directly: UO shows an approximately singly occupied U $7s$ orbital, whereas UF has an almost doubly occupied U $7s$ shell. The fractional U $6d$ population in UF again reflect metal--ligand bonding and polarization effects. 

Overall, the natural orbital analysis supports the following dominant uranium-centered valence configurations for the neutral diatomics: $5f^27s^1$ for UC and UN, $5f^37s^1$ for UO, and $5f^37s^2$ for UF (Table~\ref{tab:no_analysis}). The apparent deviations from these simple assignments in the Mulliken populations arise naturally from covalency, polarization, and the mixed U $5f/6d$ character of the bonding orbitals.

\begin{table*}[t]
\centering
\caption{Adiabatic ionization potential (AIP, in eV) for Uranium based diatomic molecules at sfX2C and X2C levels of theory, compared with reference values.}
\label{tab:aip}
\setlength{\tabcolsep}{5pt}
\renewcommand{\arraystretch}{1.2}
\begin{tabular}{l l ccc ccc l}
\toprule
\multirow{2}{*}{System} & \multirow{2}{*}{Method} 
 & \multicolumn{3}{c}{sfX2C} & \multicolumn{3}{c}{X2C} & \multirow{2}{*}{Reference} \\
\cmidrule(lr){3-5} \cmidrule(lr){6-8}
 & & TZ & QZ & $\infty$Z & TZ & QZ & $\infty$Z &  \\
\midrule
\multirow{3}{*}{UC}
 & HF     & 5.759 & 5.762 & 5.763 & 5.588 & 5.541 & 5.513 & \\
 & sc$GW$ & 6.371 & 6.405 & 6.425 & 6.227 & 6.221 & 6.218 & \\
 & FPD\cite{de_melo_theoretical_2022} & & & & & & & 6.343 \\
\midrule
\multirow{5}{*}{UN}
 & HF     & 5.692 & 5.696 & 5.698 & 5.608 & 5.579 & 5.562 & \\
 & sc$GW$ & 6.322 & 6.361 & 6.384 & 6.240 & 6.244 & 6.246 & \\
 & FPD\cite{battey_spectroscopic_2020} & & & & & & & 6.301 \\
 & SO-CASPT2\cite{battey_spectroscopic_2020} & & & & & & & 6.218 \\
 & Expt.\cite{battey_spectroscopic_2020} & & & & & & & 6.2987(3) \\
\midrule
\multirow{5}{*}{UO}
 & HF     & 5.343 & 5.347 & 5.349 & 5.295 & 5.311 & 5.320 & \\
 & sc$GW$ & 6.075 & 6.099 & 6.113 & 5.980 & 6.035 & 6.066 & \\
 & FPD\cite{romeu_energetic_2024} & & & & & & & 5.976 \\
 & X2CAMF-CCSD(T)\cite{zhang_route_2022} & & & & & & & 5.999 \\
 & Expt.\cite{han_accurate_2003,han2004electronic} & & & & & & & 6.0313 \\
\midrule
\multirow{5}{*}{UF}
 & HF     & 5.165 & 5.168 & 5.170 & 5.231 & 5.232 & 5.233 & \\
 & sc$GW$ & 6.053 & 6.087 & 6.107 & 6.100 & 6.151 & 6.180 & \\
 & FPD\cite{romeu_energetic_2024} & & & & & & & 6.278 \\
 & SO-CASPT2\cite{bross_theoretical_2015} & & & & & & & 6.337 \\
 & Expt.\cite{antonov_spectroscopic_2013} & & & & & & & 6.34159(6) \\
\bottomrule
\end{tabular}
\end{table*}

\subsection{Adiabatic IP}\label{sec:aip}

In this section, we analyze the adiabatic ionization energies (AIP) of UC, UN, UO and UF while employing the sc$GW$ approach at the scalar relativistic level (sfX2C) as well as two-component level (X2C).
The comparison between these two relativistic treatments allows for an assessment of the importance of spin-orbit coupling effects on AIP.
We also investigate the basis-set convergence for these systems at the sc$GW$ level.
AIP is evaluated when molecule in the process of either ionizing or attaching an electron is undergoing a geometry relaxation/change. This means that, in order to evaluate these quantities, the total energy of the neutral molecule and the cation or anion species with relaxed geometry have to be calculated. 
In this context, it is worth noting that adiabatic nuclear gradients within the $GW$ framework are now being developed and are available at least in the single-shot $GW$ limit, thereby enabling geometry optimizations of charged states directly within a $GW$-based formalism.~\cite{kitsaras2026analytic}

We would like to emphasize that such a procedure stands in stark contrast to traditionally performed evaluations of the vertical IPs and EAs that are conducted in the $GW$ approaches by examining the poles of the one-body Green's function without the use of the total energy. In these traditional approaches, geometry relaxation is not taken into account, and consequently, they do not illustrate AIP experiments well.
%We note that this distinction becomes less critical in the limit of a fully self-consistent $GW$ treatment. At self-consistency, the total energy is well-defined and stationary with respect to the Green's function, and the adiabatic and vertical quantities are expected to converge as the description of the potential energy surfaces of both the neutral and charged species becomes internally consistent.

In Table \ref{tab:aip}, we present AIP values obtained from sc$GW$ at the X2C and spin-free X2C levels.
We performed calculations with triple- and quadruple-zeta basis sets and subsequently extrapolated the results.
The sc$GW$ method, similar to other correlated methods, is sensitive to the size of the basis set and a better estimate for ionization energy is obtained at the extrapolated limit. For these compounds, to enable comparisons, we also present HF results, experimental estimates (when they are available), as well as best theoretical data available. 

HF calculations consistently underestimate the IPs across all four systems investigated, with the smallest deviation from experiment being 0.7 eV for UN.
At the sc$GW$ level, the results are much closer to high accuracy computational estimate and experimental values.

Comparing the different treatments of relativity shows that spin-free scalar relativistic corrections (sfX2C) already provide qualitatively correct results for these systems at the sc$GW$ level. However, sfX2C-sc$GW$ generally overestimates the adiabatic ionization potentials (AIPs) of UC, UN, and UO. Including spin–orbit coupling through the two-component X2C framework reduces this overestimation and yields improved AIP predictions for these molecules.
In particular, UO and UN show excellent agreement with experiment, with errors below 0.1 eV in both cases. 
It is worth noting that in case of UN, the sc$GW$ values are in better agreement than the state of the art spin-orbit CASPT2 computed in Ref. \citen{battey_spectroscopic_2020} at much lower computational cost. 
At the complete basis set (CBS) limit, sfX2C-sc$GW$ AIPs differ from the UC FPD reference and the experimental values for UN, UO, and UF by about 0.08–0.23 eV, with the largest deviation found for UF. By contrast, X2C-sc$GW$ reduces these deviations to about 0.03–0.16 eV.

\begin{table*}[t]
\centering
\caption{Adiabatic electron affinity (AEA, in eV) at the sfX2C level with standard and augmented basis sets, compared with reference values.}
\label{tab:aea}
\setlength{\tabcolsep}{5pt}
\renewcommand{\arraystretch}{1.2}
\begin{tabular}{l l ccc ccc l}
\toprule
\multirow{2}{*}{System} & \multirow{2}{*}{Method}
 & \multicolumn{3}{c}{cc-pV$n$Z (sfX2C)}
 & \multicolumn{3}{c}{aug-cc-pV$n$Z (sfX2C)}
 & \multirow{2}{*}{Reference} \\
\cmidrule(lr){3-5} \cmidrule(lr){6-8}
 & & TZ & QZ & $\infty$Z & TZ & QZ & $\infty$Z & \\
\midrule
\multirow{3}{*}{UC}
 & HF     & 0.471 & 0.486 & 0.495 & 0.484 & 0.491 & 0.495 & \\
 & sc$GW$ & 1.110 & 1.152 & 1.176 & 1.191 & 1.231 & 1.254 & \\
 & FPD\cite{de_melo_theoretical_2022} & & & & & & & 1.493 \\
\midrule
\multirow{3}{*}{UN}
 & HF     & 0.425 & 0.439 & 0.447 & 0.435 & 0.440 & 0.442 & \\
 & sc$GW$ & 1.139 & 1.179 & 1.201 & 1.184 & 1.210 & 1.225 & \\
 & FPD\cite{de_melo_electronic_2022} & & & & & & & 1.402$^a$ \\
\midrule
\multirow{4}{*}{UO}
 & HF     & 0.160 & 0.177 & 0.187 & 0.171 & 0.179 & 0.184 & \\
 & sc$GW$ & 0.834 & 0.877 & 0.902 & 0.877 & 0.906 & 0.923 & \\
 & FPD\cite{romeu_energetic_2024} & & & & & & & 1.123$^a$ \\
 & Expt.\cite{czekner2014high} & & & & & & & 1.1407 \\
\bottomrule
\end{tabular}
\\
\smallskip
\noindent\footnotesize $^a$FPD reference values for UN and UO were obtained at slightly 
different geometries than those used in this work.
\end{table*}

UF stands out as the most challenging system among those studied, exhibiting markedly different behavior from UC, UN, and UO. For UF, spin-orbit coupling plays a more significant role: the inclusion of X2C notably improves the AIP relative to the sfX2C result, bringing it closer to the experimental value. 

\begin{table*}[t]
\centering
\caption{Vertical Detachment Energy (VDE, in eV) at the sfX2C level with standard and augmented basis sets, compared with reference values.}
\label{tab:vea}
\setlength{\tabcolsep}{5pt}
\renewcommand{\arraystretch}{1.2}
\begin{tabular}{l l ccc ccc l}
\toprule
\multirow{2}{*}{System} & \multirow{2}{*}{Method}
 & \multicolumn{3}{c}{cc-pV$n$Z (sfX2C)}
 & \multicolumn{3}{c}{aug-cc-pV$n$Z (sfX2C)}
 & \multirow{2}{*}{Reference} \\
\cmidrule(lr){3-5} \cmidrule(lr){6-8}
 & & TZ & QZ & $\infty$Z & TZ & QZ & $\infty$Z & \\
\midrule
\multirow{4}{*}{UC}
 & HF     & 0.427 & 0.442 & 0.451 & 0.477 & 0.485 & 0.489 & \\
 & sc$GW$ & 1.107 & 1.152 & 1.179 & 1.216 & 1.256 & 1.279 & \\
 & FPD\cite{de_melo_theoretical_2022} & & & & & & & 1.487 $\pm$ 0.035 \\
 & Expt.\cite{de_melo_theoretical_2022} & & & & & & & 1.500 \\
\midrule
\multirow{4}{*}{UN}
 & HF     & 0.455 & 0.469 & 0.477 & 0.506 & 0.512 & 0.515 & \\
 & sc$GW$ & 1.176 & 1.224 & 1.252 & 1.260 & 1.290 & 1.307 & \\
 & FPD\cite{de_melo_electronic_2022} & & & & & & & 1.423 \\
 & Expt.\cite{de_melo_electronic_2022} & & & & & & & 1.424 \\
\midrule
\multirow{3}{*}{UO}
 & HF     & 0.140 & 0.158 & 0.168 & 0.209 & 0.218 & 0.224 & \\
 & sc$GW$ & 0.844 & 0.897 & 0.928 & 0.933 & 0.968 & 0.989 & \\
 & FPD\cite{romeu_energetic_2024} & & & & & & & 1.140 \\
\bottomrule
\end{tabular}
\end{table*}

\subsection{Adiabatic Electron Affinities and Vertical Detachment Energies}\label{sec:aip}

Adiabatic electron affinities (AEA) and vertical detachment energies (VDE) quantify the stability of the anionic state and provide complementary measures of electron attachment and detachment. The AEA is defined as the energy difference between the neutral and anion at their respective relaxed equilibrium geometries, whereas the VDE probes electron detachment at the anion equilibrium geometry without structural relaxation. Experimentally, VDEs are commonly obtained from anion photoelectron spectroscopy by photodetaching electrons from mass-selected anions and analyzing their kinetic energies.

\begin{equation}
\text{VDE} = E_{\text{neutral}}(R_{\text{anion}}) - E_{\text{anion}}(R_{\text{anion}}).
\end{equation}

We investigate AEA and VDE at the sc$GW$ level, examining basis set requirements for UC, UN, and UO, and the critical role of spin-orbit coupling for UF. Comparisons are made with composite FPD, CCSD and available experimental values.

\subsubsection{UC, UN and UO: Basis Set Effects}
UC, UN, and UO serve as a natural starting point for assessing electron attachment in uranium diatomics, because the added electron primarily occupies a diffuse, largely U $7s$-like orbital. 
Hence accurate prediction of electron affinities requires careful attention to basis set selection, as the attached electron in anionic states is often weakly bound and spatially diffuse.\cite{van2015, Ren2012} 
This is a well-known challenge across heavy-element systems more 
broadly: basis sets optimized for neutral atoms can be inadequate or 
even qualitatively incorrect for anionic states, as demonstrated for 
oganesson (Og), where traditional basis sets yield an EA of the wrong 
sign and reoptimization is required.\cite{Kaygorodov2021} 
To assess basis-set requirements for actinide systems, we investigated the effect of basis-set choice on the adiabatic electron affinities  of UC, UN, and UO. As shown in Table~\ref{tab:aea}, calculations were performed at the sfX2C level at the triple-zeta (TZ) and quadruple-zeta (QZ) levels, with extrapolation to the complete basis set ($\infty$Z) limit. The uranium atom was described using cc-pV$n$Z-DKH3 ($n$ = T, Q) throughout, while the ligand atoms employed either cc-pV$n$Z (standard) or aug-cc-pV$n$Z (augmented) basis sets. 
The effect of X2C is fairly minimal in these systems and we present those results in Supporting Information.

From the HF results, inclusion of augmented basis sets leads to only a 
modest increase in the AEA across all systems. The small 
magnitude of this effect at the HF level reflects the absence of 
electron correlation, which plays a more significant role in 
determining the AEA.

The sc$GW$ results show a more pronounced impact when using augmented basis sets as shown in Table~\ref{tab:aea}. 
For all systems, the EA systematically increases when using augmented basis sets. 
This highlights that for correlated methods like sc$GW$, the inclusion of diffuse functions is more critical, as these functions allow a better description of the electron correlation in the anionic state.

The sc$GW$ method systematically underestimates the EA relative to the FPD composite approach. 
This systematic underestimation is expected and arises primarily from the absence of higher-order correlation effects in the $GW$ approximation. The FPD method includes connected triple excitations through CCSD(T), full triples via CCSDT, and partial quadruple excitations through CCSDT(Q), which collectively contribute an additional 0.10--0.15 eV to the EA.\cite{de_melo_theoretical_2022,de_melo_electronic_2022} Additional FPD corrections include scalar relativistic effects beyond sfX2C, QED contributions (Lamb shift), and the Gaunt term accounting for spin-other-orbit coupling. 
Despite these systematic differences, the sc$GW$ method captures the essential physics of electron attachment and provides a computationally efficient alternative for screening electron affinities in actinide systems, with errors that are predictable and well-characterized. 
We note that the geometries used for the FPD calculations differ slightly from those used in our $GW$ calculations, which may contribute minor differences in the comparison, though the systematic trends remain consistent. This underestimation is consistent with previous $GW$ studies.~\cite{van2015, Ren2012}

The results for VDE are reported in Table~\ref{tab:vea}. Several clear trends emerge from the basis set study. First, sc$GW$ consistently predicts larger VDEs than HF across all systems, reflecting the inclusion of electron correlation effects that stabilize the anionic state. Second, the use of augmented basis sets systematically increases VDEs, particularly at the triple-zeta level, highlighting the importance of diffuse functions for describing the spatially extended extra electron.

Comparison with experimental and FPD reference values shows that sc$GW$ provides significant improvement over HF but still systematically underestimates the VDE. The VDE underestimation by sc$GW$ is similar in magnitude to that observed for the AEA (0.18--0.24 eV), indicating that the missing higher-order correlation effects contribute comparably to both vertical and adiabatic electron detachment processes. The consistency of these systematic errors across the actinide monocarbide, mononitride, and monoxide series demonstrates that sc$GW$ provides a reliable and computationally efficient method for predicting electron detachment energies, with well-characterized and predictable deviations from benchmark values.

\begin{figure}
  \centering
  \includegraphics[width=.85\linewidth]{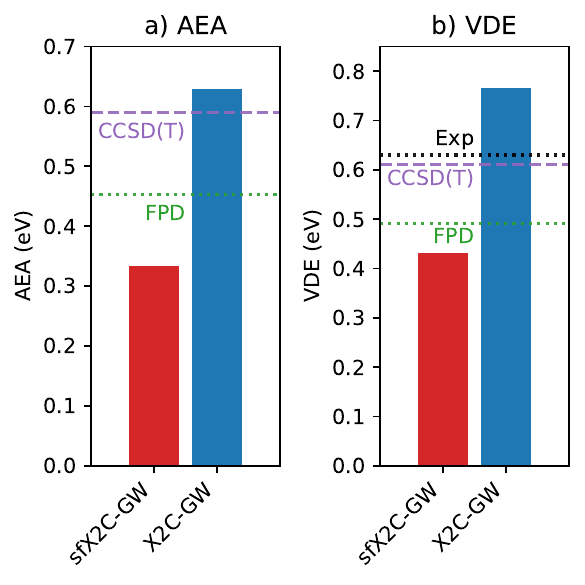}
  \caption{ a) AEA and b) VDE of the UF molecule calculated at the sc$GW$ level using two relativistic treatments, sfX2C and X2C, with the QZ basis. Reference values from FPD ~\cite{romeu_energetic_2024}, X2C-CCSD(T)~\cite{burak2024}, and experiment~\cite{burak2024} are also shown.}
  \label{fig:soc_uf}
\end{figure}

\begin{table*}[t]
\centering
\caption{Vibrational frequency ($\omega_e$, in cm$^{-1}$) and bond length ($r_e$, in \AA) for diatomic uranium molecules using the scalar relativistic sc$GW$ approach, compared with other methods and experiment.}
\label{tab:vib_freq}
\setlength{\tabcolsep}{6pt}
\renewcommand{\arraystretch}{1.2}
\begin{tabular}{l l c c}
\toprule
Molecule & Method & $\omega_e$ & $r_e$ \\
\midrule
\multirow{6}{*}{UC}
 & sc$GW$ & 920.35 & 1.843 \\
 %& SO-CASPT2 (4e,9o)\cite{de_melo_theoretical_2022} & 949.6 & 1.854 \\
 & CCSD(T)\cite{de_melo_theoretical_2022} & 922.0 & 1.8699 \\
 %& SO-CASPT2 (10e,12o)\cite{wang2011infrared} & 945 & 1.863 \\
 & SO-CASPT2 (10e,16o)\cite{pogany2016theoretical} & 928 & 1.870 \\
 & Expt.\cite{wang2011infrared} & 827.8$^a$, 871.7$^b$ & \\
\midrule
\multirow{5}{*}{UN}
 & sc$GW$ & 1026.2 & 1.733 \\
 & CCSD(T)\cite{de_melo_electronic_2022} & 1017.8 & 1.762 \\
 & SO-CASPT2 (3e,8o)\cite{de_melo_electronic_2022} & 1006 & 1.768 \\
 %& SO-CASPT2 (3e,8o)\cite{battey_spectroscopic_2020} & 1010 & 1.766 \\
 & Expt.\cite{green1976identification,matthew_resonant_2013} & 996$^c$ & 1.7650(12) \\
\midrule
\multirow{4}{*}{UO}
 & sc$GW$ & 825.53 & 1.822 \\
 & CCSD(T)\cite{romeu_energetic_2024} & 849.63 & 1.8379 \\
 & B3LYP\cite{han2022electronic} & 846.3 & 1.847 \\
 & SO-CASPT2 (6e,9o)\cite{romeu_energetic_2024} & 854.76 & 1.8444 \\
 & Expt.\cite{kaledin_laser_1994} & 846.5 & 1.8383 \\
\midrule
\multirow{5}{*}{UF}
 & sc$GW$ & 553.33 & 2.022 \\
 & CCSD(T)\cite{romeu_energetic_2024} & 577.99 & 2.0295 \\
 & SO-CASPT2 (5e,13o)\cite{romeu_energetic_2024} & 582.83 & 2.0280 \\
 & B3LYP\cite{vent2015reactions} & 573.3 & 2.037 \\
 & Expt.\cite{vent2015reactions,antonov_spectroscopic_2013} & 567.7 & 2.020 \\
\bottomrule
\end{tabular}
\smallskip
\\
\noindent{\footnotesize $^a$ Ar matrix, $^b$ Ne matrix, $^c$ Ar matrix }
\end{table*}

\subsubsection{UF: Effect of Spin-Orbit Coupling}

UF represents a particularly interesting test case for evaluating the treatment of spin-orbit coupling in actinide systems. 
Upon electron attachment to form UF$^-$, since the 7$s$ is already doubly occupied, the additional electron occupies a uranium 6d orbital (U$^+$[5f$^3$7s$^2$] + e$^-$ $\rightarrow$ U[5f$^3$6d$^1$7s$^2$]), where strong SOC effects in the 6d shell significantly influence the electronic structure.\cite{burak2024} Recent benchmark X2C-CCSD(T) calculations have established accurate reference values for this system: VDE = 0.61 eV and AEA = 0.59 eV, in excellent agreement with the experimental VDE of 0.630(30) eV.\cite{burak2024} Notably, composite FPD methods that treat scalar relativistic effects variationally but SOC perturbatively yield VDE = 0.492 eV and AEA = 0.453 eV which is significantly below experimental values, contrasting with their success for UC and UN.\cite{romeu_energetic_2024} This behavior highlights UF as a critical benchmark for methods treating relativistic effects in actinide chemistry where varitational inclusion of SOC effects become important.

We present the $GW$ results for UF in Figure~\ref{fig:soc_uf}. The sfX2C-$GW$ approach, which neglects SOC, significantly underestimates both the VDE (0.432 eV) and AEA (0.334 eV)—errors of 0.20 eV and 0.26 eV relative to experiment, respectively. The X2C-$GW$ method, which incorporates SOC directly into the spinor orbitals, yields substantially improved values of 0.765 eV and 0.629 eV for the VDE and AEA. While X2C-$GW$ overshoots the experimental VDE by 0.14 eV, this represents a marked improvement over both sfX2C-$GW$ and the FPD approach.

The remaining X2C-sc$GW$ discrepancy likely reflects three sources of error: missing higher-order correlation beyond the $GW$ approximation, incomplete basis-set convergence, and residual relativistic errors from the approximate treatment of two-electron picture-change effects in the X2C Hamiltonian. The benchmark calculations employed basis-set extrapolation to approach the complete-basis-set limit,\cite{burak2024} and specialized contraction schemes for two-component basis sets\cite{zhang_new_2024} may help reduce basis-set errors in future $GW$ implementations. Comparisons with atomic-mean-field X2C treatments can further clarify the role of two-electron picture-change effects in the remaining discrepancies.
The inclusion of two-electron contribution at the X2C-AMF level at a moderate basis set is discussed in the Supporting information.

These results demonstrate that when SOC plays a decisive role in determining orbital occupancy upon electron attachment—particularly for actinide d or f orbitals—variational treatment of SOC is essential for quantitative accuracy. The X2C-$GW$ approach provides significant improvement over scalar relativistic methods and represents a computationally efficient alternative for screening studies of actinide systems where benchmark coupled-cluster calculations may be prohibitively expensive.

\subsection{Bond length and Frequency}

In this section, we investigate the equilibrium bond length and vibrational frequency using our sc$GW$ method and compared with existing high accuracy computational chemistry methods and experimentally estimated values.
Calculations of bond lengths and harmonic frequencies for the diatomic molecules were carried out using the cc-PVTZ basis set for ligands: C, N, O and F \cite{dunning_gaussian_1989}, while the U maintained the cc-pvTZ-DKH3 basis used from the energetic calculations shown above \cite{peterson_correlation_2015}. 
The potential energy curves were explored by computing energies at 5 equidistant points, spread 0.02$\AA$ from the targeted equilibrium geometry with additional 0.04 $\AA$ to form a well-rounded curve. 
The details of these can be found in the supporting information.
An 5th-order polynomial was fitted to the resulting energy data, and the equilibrium bond lengths and harmonic frequencies were subsequently determined from the polynomial coefficients.
%We do not include any other corrections on top of the result like zero-point correction or SOC effects for this section, and hence our estimates are only a reference. 

In Table \ref{tab:vib_freq}, we present the equilibrium bond length ($r_e$) and vibrational frequency $\omega_e$ for the Uranium molecules we considered in the study. 
For all the systems, we can generally see that sc$GW$ gives very good estimate for both vibrational frequency and bond-lengths although bond-length is slightly underestimated at the sc$GW$ level. 
This can be mainly due to the exchange and correlation not included in equal footing and has been seen in earlier studies as well.\cite{abraham_relativistic_2024}

For UC, there is no gas phase experimental data and the values using Neon and Argon matrix vary quite much suggesting strong interaction between the matrix and UC molecules. 
Hence, we compare with SO-CASPT2 and CCSD(T) values. 
%It has to be noted that the SO-CASPT2 as well as CCSD(T) values are corrected for  from zero point correction {\color{red}FILL} which is not included in this case.
The estimated vibrational frequency using sc$GW$ is very close to the CCSD(T) estimate. 
For UN again gas phase estimate of vibrational frequency is not available and the sc$GW$ values are within 10 cm$^{-1}$ compared to the highly accurate CCSD(T) estimate.
In Table~\ref{tab:vib_freq}, we report only the most recent estimate from Romeu \emph{et al.}, who also provide a comprehensive summary of previous estimates.\cite{romeu_energetic_2024}
Except for UF, sc$GW$ consistently underestimates bond lengths. For UF, however, it yields notably accurate bond length.

\section{\label{sec:conc} Conclusions}

This work presents, to our knowledge, the first application of an all electron fully self-consistent $GW$ method to actinide molecules. Using UC, UN, UO, and UF as benchmarks, we show that all-electron X2C-sc$GW$ can be converged for challenging open-shell uranium compounds and provides a practical, starting-point--independent framework for their energetics and spectroscopy. Across the series, sc$GW$ yields accurate adiabatic ionization energies and good predictions of equilibrium bond lengths and harmonic vibrational frequencies.

For AEA and VDE values, diffuse basis augmentation is essential to describe the attached electron and obtain converged results. UF provides the clearest example of the additional role of relativity: upon electron attachment, the added electron occupies a uranium $6d$ orbital, making the attachment and detachment energies strongly sensitive to spin--orbit coupling. As a result, UF requires a variational two-component X2C treatment of spin--orbit coupling for reliable energetics.

Overall, this work highlights sc$GW$ as a powerful and reliable framework for actinide chemistry, capable of delivering accurate energetics and spectroscopic properties for systems where relativistic effects, open-shell character, and strong orbital mixing are all important. 
Two-component sc$GW$ calculations remain computationally and numerically demanding, particularly for heavy-element open-shell systems where spin--orbit coupling, near-degeneracies, and symmetry breaking can complicate convergence. 
Nevertheless, the present results demonstrate that fully self-consistent $GW$ solutions can be obtained even for challenging uranium compounds, establishing X2C-sc$GW$ as a practical approach for heavy-element molecules. 
Future developments should focus on improving both robustness and efficiency. In particular, frozen-natural-orbital-based extensions and more compact, systematically improvable basis sets will be important for reducing the cost of all-electron two-component calculations. The development of spinor-based basis sets~\cite{zhang_new_2024} provides a promising route toward more controlled basis-set convergence, while recent correlation-consistent effective core potentials with spin--orbit terms (ccECPs)~\cite{Madany2025} offer an additional strategy for extending sc$GW$ to larger actinide systems. These advances will enable applications beyond diatomic molecules, including larger actinide complexes and actinide-containing solids, where self-consistency and a balanced treatment of relativistic effects are expected to be especially important.

\begin{acknowledgments}
    The authors would like to thank Prof. Lan Cheng, Dr. Xubo Wang, and Dr. Chaoqun Zhang for their valuable discussions.
    The authors would also like to thank Dr. Gaurav Harsha for carefully reading the manuscript and providing helpful comments. This research was supported by the U.S. Department of Energy under Award No. DE-SC0026306, ``Ab-initio Green's function methods describing relativistic effects.''
\end{acknowledgments}

\bibliographystyle{achemso}
\bibliography{bibfile}% Produces the bibliography via BibTeX.

\end{document}